# Fast Neutron Resonance Radiography in a Pulsed Neutron Beam

V. Dangendorf[+], G. Laczko[+], C. Kersten[+], O. Jagutzki[*], U. Spillmann[*]
[+]Physikalisch-Technische Bundesanstalt (PTB), Braunschweig, Germany
*Universität Frankfurt, Frankfurt/M, Germany

*Abstract*— The feasibility of performing fast neutron resonance radiography at the PTB accelerator facility is studied. A neutron beam of a broad spectral distribution is produced by a pulsed 13 MeV deuterium beam hitting a thick Be target. The potential of 3 different neutron imaging detectors with time-of flight capability is investigated. The applied methods comprise wire chambers with hydrogenous converter layers and a fast plastic scintillator with different optical readout schemes.

We present the neutron facility, the imaging methods employed and results obtained. in beam experiments where samples of carbon rods with various length and diameter were imaged to study resolution and sensitivity of the method.

I. INTRODUCTION

Fast neutron resonance radiography is one of the suggested methods to measure the 2-dimensional elemental distribution of low-Z elements in samples. Fast neutrons can penetrate massive samples and their interaction cross section shows comparably small variation between the different elements, (compared to X-rays or thermal neutrons). Possible applications are in air port security (luggage and cargo inspection [1,2], mining [3] and in the field of NDT, (non destructive testing) where sensitivity to the distribution of light elements in the presence of high Z shielding material is required [4]. The method is based on the distinct energy dependence of the neutron cross section of the elements of interest. By taking several radiographs with neutrons of selected energies it is possible to derive the spatial distributions of the wanted elements either by digital subtraction of images if the distribution of single elements is wanted [3]) or by unfolding methods if mapping of the distribution of several elements is of interest [1,2].

For digital subtraction radiography with two distinct energies powerful methods with intensive, almost monoenergetic neutron beams produced with high current accelerators and pressurised gas targets are developed [3,4]. Energy variation is obtained by switching of the last stage of a linac [3,5] or by changing the viewing angle of detector/sample to the target [1,2]. Due to the high neutron yield the imaging detector must be an integrating system, typically scintillators with CCD-camera read out. Fast time resolution is not required because each frame is exposed with one energy and images of different energies are acquired one after the other.

Alternatively, thick targets for neutron production can be used, where almost a "white" neutron energy distribution is obtained in a specific range, e.g. from 1 MeV up to 10 MeV. Energy selection is realised by the Time-Of-Flight (TOF) method. Neutron beams with a broad spectral distribution can be produced by bombarding thick metal or liquid targets of Be or Li. Compared with the previously described method the neutron yield per unit beam charge is much higher, therefore the accelerators need to deliver only tens of microampere instead of several milliampere in ion current. On the other hand, due to the application of TOF methods the beam has to be delivered in short, nanosecond wide pulses with pulse separation times of several 100 ns. Furthermore, the TOF technique requires neutron imaging techniques with a time resolution of the order of a few ns. Due to the availability of intense pulsed neutron beams between 1 and 15 MeV at the PTB-accelerator facility [6] and detection and imaging equipment capable for TOF measurement, it was decided to investigate the possibilities at PTB to perform fast neutron radiography with TOF methods.

II. THE EXPERIMENT

*A. Neutron Production and Time-of-Flight Facility*

The PTB accelerator facility for fast neutron research is described in detail in [7]. Neutrons are produced by accelerated light ions and collision with selected target nuclei to produce defined monochromatic neutron fields as well as neutrons of a broad spectral distribution. The ion accelerators are a rotatable energy variable cyclotron (TCC CV28 at positon 1 in Fig 1) and a 3,75 MeV Van-de-Graff accelerator. Both accelerators are able to provide pulsed beams with ca 1 ns pulse width for TOF-applications. At several lines "standard" neutron and photon fields for calibration of dosemeters and metrological investigations are provided. A "high current" beam line (see 2 in fig 1) is available with a collimated neutron beam, originally designed for irradiating biological samples and technical materials and for developing instrumentation for neutron therapy. Recently also a microbeam facility was installed where controlled irradiation of material probes and biological cells with single ions at a precision of around one micrometer is feasible. Close to the cyclotron a target for neutron production can be inserted at the point of the cyclotrons rotation center (see pos.3 in fig 1). By turning the Cyclotron this technique allows for variation of the angle between the beam axis and the sample / detector axis. This station is used for measuring double differential neutron scattering cross sections of various materials relevant to nuclear technology.

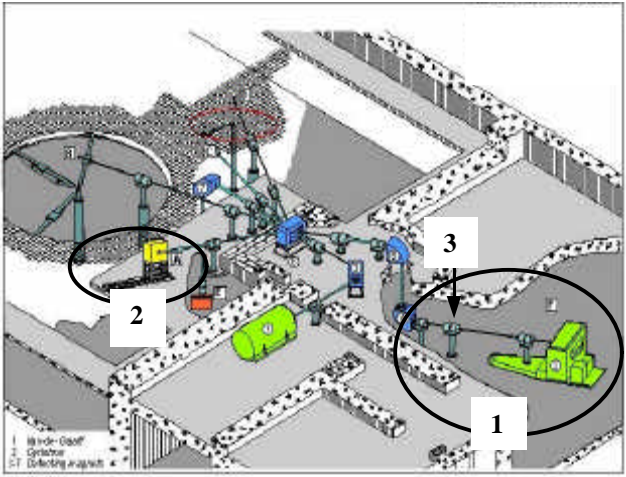

*Fig.1: Neutron research facility at PTB. (1) is the energy variable and rotatable cyclotron , (2) the „high current" neutron irradiation facility with the collimator and (3) marks the place, where optionally a neutron producing target can be installed to enable measurements at variable angles.*

A 3 mm thick Be-Target in the center of a massive collimator is bombarded with deuterons of 13 MeV energy. The pulse separation time is 500 ns. Optionally this interval could be increased up to about 40 µs (with loss of beam current). The pulsing method employed at PTB is the TCC internal pulsing system described in [7,8] which provides single turn extraction of one beam pulse with high efficiency and good time resolution. The time resolution was determined by measuring the width of the time distribution of the γ-flash with a Cherenkov detector. Depending on the tuning of the machine a pulse width between 1,5 and 1,8 ns (fwhm) was obtained. The available deuteron current at our experiment was 20 µA unpulsed and about 2 µA in pulsed mode. The maximum deuteron current for the cyclotron is about 100 uA.[7]. Since the pulsed injection reduces the load for the cyclotron, the limit is set by the ion source and the injection system. Presently it is discussed to overcome this limitation by an external ion source and a pulse selection scheme that injects beam pulses through the injection hole near the cyclotron center.

At present there is no direct control of the spot size of the beam on the target but from earlier runs and visual inspection of the target it was concluded that the diameter varies between 1 and 3 mm. A beam profile measurement close to the target is not feasible due to lack of space inside the collimator. Setting up the whole experiment in the cyclotron cave close to the cyclotrons rotation point (position (3) in fig 1) would overcome this limitation. At this place the short beam transport to the target is expected to improve also the time structure of the beam.

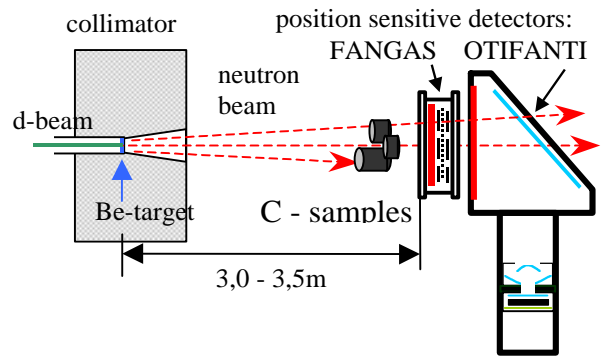

*Fig.2: Experimental setup of the fast neutron radiography experiment. The neutrons are produced by 13 MeV deuterons hitting a 3 mm thick Be target inside a massive steel and lead collimator. The sample is assembled of carbon rods (see fig. 4). Several types of detectors (FANGAS and OTIFANTI) are used in the experiments.*

### B. Experimental Setup

The radiography experiment was installed at the "high current" station of the PTB neutron facility (see fig 2 position 2). A schematic drawing is shown in fig. 2. Inside the collimator neutrons are produced by the 13 MeV Be(d.n) reaction. The collimator forms a neutron beam with a square beam profile and a divergence of about 6,5°, thus at the place of the detectors the neutron field covers a field of about 70 * 70 cm$^2$. The collimation of the beam is required to reduce radiation protection problems in the experimental hall and to avoid damage of electronics and the electro-optical elements of the detectors, which are placed outside the beam. The distance between the sample and the neutron source was of the order of 3 m. The distance between the sample and the detectors is varied between 50 and 900 mm to study the influence of scattered neutrons from the sample.

For this irradiation geometry the neutron flux through the detectors is calculated by the data of Brede et al [6]. Fig. 3 shows the forward neutron yield per unit beam charge. For a typical pulsed deuteron beam of 2 µA a neutron fluence rate at a detector position 3 m away from the target of $3*10^5$ s$^{-1}$cm$^{-2}$ is obtained.

The samples consist of assemblies made of rods of graphite of various length (the extension in beam direction) and diameter to study the sensitivity and the position resolution. Fig 4 shows the two sets of samples used in these experiments. Carbon is chosen because it has a broad resonance structure between 6,5 and 8,2 MeV (see Fig. 5). The two shaded fields in this graph show two broad energy regions which were selected by appropriated TOF–conditions to generate two images with neutrons of energies at strongly different values of the carbon cross section in order to study the feasibility of the resonance radiography method.

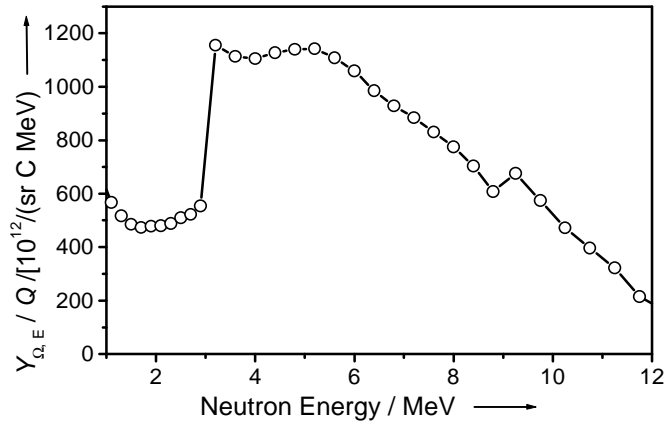

*Fig. 3: Neutron yield in forward direction of 13 MeV deuterons hitting a thick Be target [6].*

Three different types of detectors are studied for their applicability in these kind of experiments. FANGAS (fig.6) is a wire chamber with polyethylene converter, OTIFANTI (fig. 7) a fast scintillator, read out alternatively either by a quantum counting photon imaging system (PCII) or a high speed fast framing camera. In the following the detection methods will be described.

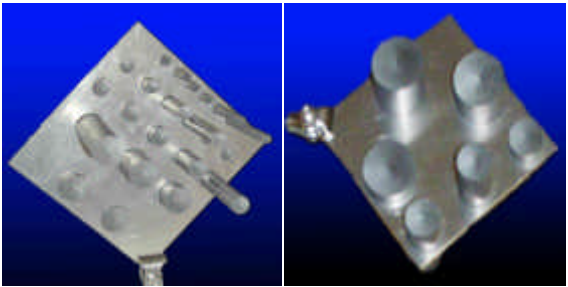

*Fig.4: The samples made of carbon rods. The diameters range from 5 mm to 30 mm, the length (which extends parallel to the*

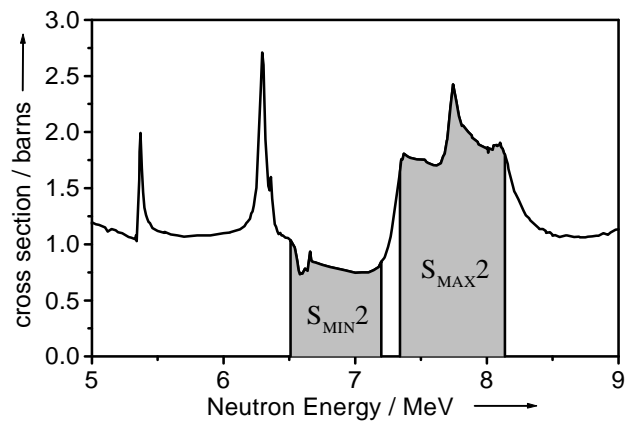

*neutron beam axis) from 5 to 60 mm.*

*Fig.5. Total neutron cross section of carbon in the energy range 5 – 9 MeV. The marked regions $S_{MIN}2$ and $S_{MAX}2$ show the energy windows used for studying the digital subtraction method. Data are taken from ENDF/B6 [9]*

### C. Detectors

FANGAS, the FAst NeutronGAS-filled imaging detector is based on an earlier development of a thermal neutron detector with a $^6$Li or Gd converter coupled to a multi-step wire-chamber [10]. Instead of the original thermal neutron converter a polyethylene (PE) radiator, 1 mm in thickness, is chosen for fast neutron to proton conversion. Fig. 6 shows a schematic drawing of the detector. Neutrons interact in the PE with hydrogen protons by elastic scattering. Recoil protons which escape from foil ionize the gas along the track. Electrons from the region close to the foil surface are multiplied in the first step within a Parallel Plate Avalanche Chamber (PPAC). Final amplification and localisation takes place in a Multi-Wire Proportional Chamber (MWPC). Position coding is obtained by cathode delay line readout. This detector has a sensitive area of 314 cm$^2$ (20 cm in diameter).

The data acquisition system is based on the Roentdek HM1 module with the COBOLD data acquisition system. [11]. Due to the need of high resolution TOF-information associated with the position coordinates of each detected neutron list-mode data storage is applied, which maintains the correlation of all relevant parameters for each event. This limits the acquisition speed to about $2*10^4$ events per second. The system offers alternatively also a 2d-histogramming mode with a selected number of TOF channels which is able to handle up to $10^6$ s$^{-1}$.

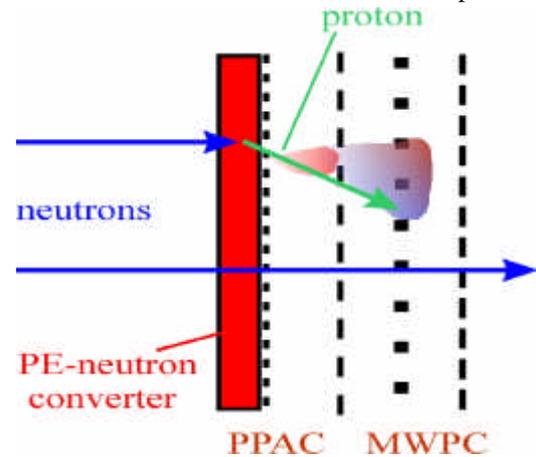

*Fig. 6: Schematic view of the FAst Neutron GAS detector (FANGAS)*

Polyethylene is chosen as radiator due to its large hydrogen content and ease of use. The polyethylene is sandwiched between two stretched stainless steel meshes which support the radiator and act as cathode of the PPAD. The thickness of the radiator is 1 mm, which is the optimal compromise between the demand of a thick converter for large neutron absorption and the limited escape length for protons from the PE. The calculation of the radiator efficiencies was performed with the program "Radiator", developed for the simulation of proton-recoil telescope responses [12] For a 1 mm thick converter the calculated efficiency by np-scattering is 0,1 % - the carbon in the PE and contributions of other elements (windows, meshes) are not taken into account.

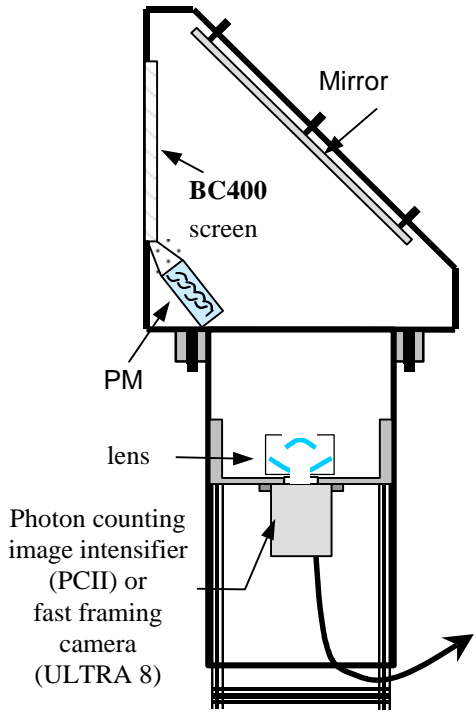

*Fig.7: Schematic view of the OpTIcal FAst NeuTron Imaging system (OTIFANTI)*

The second detector, OTIFANTI, the OpTIcal FAst NeuTron Imaging system is schematically shown in fig. 7. It consists of a square shaped fast plastic scintillator, type BC400 of 22 cm edge length and 10 mm thickness. Neutrons interact in the plastic material mainly by np-scattering. The fast proton produces a local light spot. The image of this light emission is projected via a front side coated Al mirror and a UV-mirror lens of f = 90 mm to the optical detection systems described below. A photomultiplier (PM) is connected by a perspex light guide to one edge of the scintillator to provide a fast trigger signal. Due to the good optical coupling of this PM to the scintillator this branch has almost 100 % efficiency for detecting neutrons which have interacted in the scintillator. This is used to calibrate the efficiency of the optical branch of the detector.

Two different optical image sensors were investigated in the OTIFANTI. The first one, a quantum counting image intensifier, is described in [13]. In its first stages it is a conventional Image Intensifier (II) with an S20 photocathode of 25 mm diameter and a Chevron pair of Multi- Channel Plates (MCP). The output phosphor is replaced by a layer of Ge acting as a charge collection anode. The high resistivity of this layer enables the localisation of the moving charges between MCP cathode and anode outside the vacuum tube. The readout of the position coordinates of the electron avalanche is obtained by a wire wound delay, the time pickup signal is provided either from the second MCP anode or from the Ge-layer.

The second system is a modified ULTRA8 fast framing camera from DRS Hadland [14]. The ULTRA8 is an intensified CCD-Camera wit an 9-fold segmented intensifier photocathode. Each of the segments is individually gateable. An image splitter in front of the II distributes the optical image to the 9 segments. This method allows for 8 frames (2 of the 9 frames are coupled) to be independently exposed in time intervals as short as 10 ns. The frames are collected by one large CCD with a dynamical resolution of 12 bit. Each frame covers a region of 512 *512 pixels of the CCD.

In our application we used a modified version of the standard camera, which allows for multiple exposures of the same frame at defined time delays and intervals upon receipt of a trigger. The trigger is provided by the beam pickup of the cyclotron and defines the start of the neutron TOF-measurement. The exposure time intervals have to be selected to a defined TOF window with corresponds to a preselected energy range of the neutron spectrum. In the present version of the ULTRA8 we are able to retrigger only 80 times per 2,7 ms cycle due to the limitations of the present camera software. Therefore, about 98 % of the available neutron beam is unused. This is a severe limitation of the camera in these experiments but according to the manufacturer these problem might be overcome by providing direct access to the control inputs of the high voltage pulsers of the photocathode segments.

Much more severe was the observation in our beam experiment, that no net response to the neutron beam could be obtained with the ULTRA8. This was in sharp contrast to the response of another intensified camera with extended UV response. ("OPAC" –camera, described in [15]) which replaced the ULTRA8 after its failure. The reason for the failure of the ULTRA8 is still unknown, but at present the total and relative spectral sensitivity of the ULTRA8 is investigated and compared to the data quoted by the manufacturer to assure the compatibility of the ULTRA8 and the BC400 scintillator.
In the results we present data obtained with the OPAC camera. This camera has no framing capability and is retriggerable only at a very small rate. Therefore it is not applicable for multiframe TOF experiments. But it should allow for a estimation of the expected exposure times and image quality of the ULTRA8 after overcoming its present problems.

III. RESULTS AND DISCUSSION

*A. Detection Efficiency*

FANGAS: The detection efficiency was determined by measuring the count-rate of the detector at a defined deuterium beam current. The neutron flux through the detector surface is derived from the neutron yield data of Brede et al [6]. The measured rate with FANGAS was $8*10^4$ s$^{-1}$ at a beam current of 1,5 µA. From this a spectrum averaged neutron detection efficiency of 0,19% is obtained. This is by a factor of 1,9 more than calculated. This might be due to the additional contribution of the carbon in the polyethylene and the contribution from the meshes. Still, this low detection efficiency is inadequate for a real application of this method and possible improvements have to be discussed. One option would be the stacking of many converter/amplifier modules along the beam axes. However, for reaching detection levels of a few percent stacks of several ten modules are required and the availability of cost efficient electron amplification modules is mandatory. This demand can not be satisfied by wire chambers due to their time consuming manufacture and therefore high costs. The use of GEM foils [16] which recently showed promising results in a multi layer assembly for UV–photon imaging [17] is presently investigated.

OTIFANTI: The detection efficiency of the optical detector depends firstly on the probability of a neutron induced charged particle emission in the scintillator (neutron conversion efficiency). Secondly, on the probability that from the scintillation light at least one photoelectron is produced by the photocathode of the image intensifier which is then amplified in the image intensifier (optical sensitivity). This second process depends on the efficieny of the scintillator to convert charged particle energy to photons, the geometrical solid angle covered by the imaging optics, the scatter and absorption losses of the optical components and the quantum efficiency of the image intensifier.

To selectively measure the neutron interaction efficiency, the PM attached to the scintillator screen (reference PM) was used. Similar to the procedure described above for the FANGAS the neutron conversion efficiency was measured to 2,6 %. The optical efficiency was tested by mounting a second PM ($PM_{opt}$) at the place of the image intensifiers of the optical systems and counting the coincidence rate between $PM_{opt}$ and the reference PM. It is assumed that the quantum efficiencies of the image intensifiers and $PM_{opt}$ are the same (both have glass windows, similar photocathode material and the same exposed area). The coincidence rate at the highest possible gain of $PM_{opt}$ was 60 %, which results in an overall neutron detection efficieny of 1,5 %. The dependence of the coincidence rate from the gain of $PM_{opt}$ and the form of its pulse-height spectrum showed, that the signal is produced mainly by single photo electrons.

While the neutron interaction efficiency can be improved quite easily by using thicker scintillators (though to the expense of loss in resolution due to the parallax in a divergent neutron beam and neutron scattering inside the screen) the optical sensitivity is more difficult to improve. Fast hydrogen-rich scintillators with larger light output are not available and therefore increasing the efficiency of the optics (e.g. using larger image intensifiers and lenses) have to be considered.

Data Acquisition: While the speed of the data acquisition for the CCD system is not relevant (though ultimately the duty cycle between effective exposure time and read out time might become important) the rate capability of the quantum counting devices and therefore the effective efficiency depends on the data acquisition system. For each detected neutron event 4 position coordinates and TOF have to be measured. Two position coordinates are then calculated per software. The coordinates and TOF are either stored in a list-mode file or in a 3-dimensional matrix (online histogramming). The number of time bins needed in these experiments is comparably small. Therefore, online histogramming of an appropriate number of TOF-matrixes would be the proper choice. However for this the relevant TOF bins have to be implemented at run-time. This might be suitable for a developed system. For the present development device list-mode acquisition is more appropriate due to the possibility of flexible offline data analysis. In list-mode about $2*10^4$ events per second could be acquired which is by a factor of 5 (FANGAS) up to 50 (OTIFANTI) smaller than the detected neutron rate. Therefore, for the judgement of the exposure time required for the counting detectors in this experiments, it should be kept in mind that the data acquisition is the limiting factor and not the detection efficiency or the neutron flux.

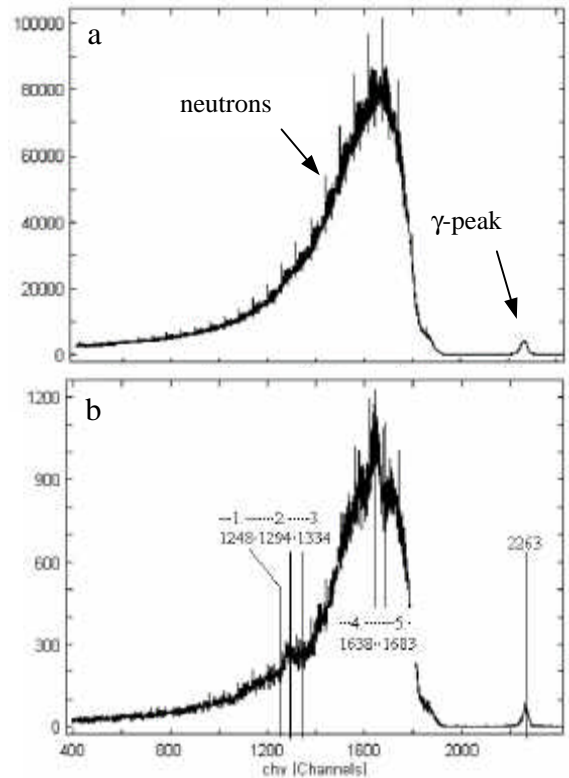

*Fig. 8: Time-of-flight spectrum of FANGAS. a) total TOF spectrum without samples in front of the detector b) behind a 60 mm column of C. For the evaluation of the detector time resolution the width of the **g**- peak was measured.*

### B. Time and Energy Resolution

Fig. 8 shows the TOF-spectrum of the FANGAS detector. The broad distribution in the central part represents the neutron spectrum in TOF scaling. The time resolution was measured by analysing the width of the γ-peak, indicated in the TOF-spectrum. The detector time resolution is obtained by quadratic subtraction of the width of the deuteron pulse from the width of the gamma peak in the TOF spectrum. For FANGAS a resolution of $\tau_{FG}$ = 4,1 ns (fwhm) and for OTIFANTI with the photon counting image intensifier PCII) $\tau_{FG}$= 3,0 ns(fwhm) is obtained. Both values are inferior to our expectations because the intrinsic resolution of the gas detector and of the image intensifier is known from earlier measurements to be < 1 ns for the IICD and about 1,5 ns for the gas detector.

To observe the effect of the energy dependent neutron cross section in carbon, the TOF spectrum behind the 60 mm carbon column was selected from the list-mode data. This spectrum is shown in fig 8b. The cross section maxima in the 3 MeV and the 7,5 – 8,2 MeV region are visible (see cursor positions) as dips in the spectrum. The energy resolution which can be calculated from the TOF resolution of FANGAS at a detector to target distance of 3,5 m is listed in Tab.2 . Also the energy resolution $\Delta E_r$ required to resolve the resonance structures is listed. While for the 8 MeV structure the time resolution satisfies the needs the resolution for the 3 MeV lines has to be improved.

| $E$ / MeV | 2 | 3 | 6 | 8 | 10 |
|---|---|---|---|---|---|
| TOF / ns | 181,3 | 148,1 | 104,7 | 90,7 | 81,1 |
| $\Delta E$ / MeV | 0,09 | 0,17 | 0,47 | 0,72 | 1,01 |
| $\Delta E_r$ / MeV | | 0,05 | | 0,6 | |

Tab.1. The table lists the TOF, the energy resolution **DE** (fwhm) with FANGAS at a target distance of 3,5 m and the required energy resolution **DE**$_r$ for the 2 carbon resonances at 3 MeV and about 8 MeV.

### C. Imaging Results

Images of satisfying quality are obtained from FANGAS and from OTIFANTI with CCD read-out. The data with OTIFANTI/PCII show only very crude and distorted position information with low contrast. This is most probable due to the bad electronic signal conditioning and the low obtainable gain of the PCII-tube which was available for our tests. The data of the measurements with FANGAS are stored in list- mode, so in the offline analysis optimisation of the TOF-window settings is possible and image corrections due to uncorrelated measurement data (random correlated X,Y and TOF-data) can be applied. Thus, for both samples images using the total neutron spectrum and 4 selected energy bins around the resonances were produced. Position dependent efficiency variations and inhomogeneities in the neutron field are corrected for by normalising the radiographic image of the samples with a reference image of the neutron beam. This reference image was obtained with the same analysis conditions but acquired without sample.

Fig. 9 shows the absorption images of the 2 samples after field and efficiency normalisation. All neutron energies were selected to contribute, only γ-events were excluded. This yields the images with the best contrast due to the comparably good statistics in each pixel. Sample thicknesses of down to 10 mm can be observed, also the samples of diameter 5 mm are seen down to 20 mm thickness. Already the images which make use of all available neutron energies are limited by Poisson noise. The energy selected images in the 8 MeV and 3 MeV region are even more limited by their counting statistics per pixel because of the narrow energy windows accepted.

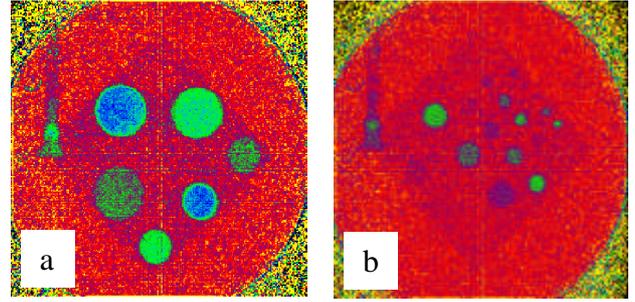

Fig. 9: Radiography with full neutron spectrum taken with the FANGAS detector, a) sample 1 (see fig.4) with 95 min exposure time and b) sample 2 with 72 min exposure time.

Subtraction of 2 images, which already suffer from excessive Poisson noise is therefore almost impossible. Nevertheless, fig. 10 shows the differential image of sample 1 in the 8 MeV region. At least the structures remain visible. With a more efficient detector and a faster data acquisition it is expected to overcome this limitation.

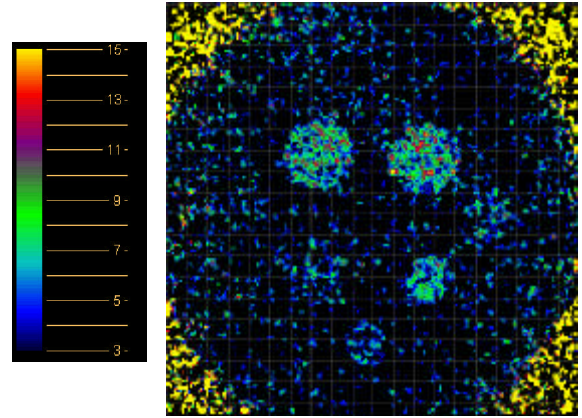

Fig 10: Digitally subtracted image of sample 1 at the 8 MeV cross section structure of carbon. The image is smoothed with a mean filter.

The OTIFANTI with CCD readout was not applicable for energy selected radiographs due to failure of the ULTRA8. Images of both samples obtained with the ungated CCD camera are shown in fig.11. The total exposure of the sample was 30 s at an ion beam current of 20 μA. This provides a neutron fluence for the exposure of this image of about $8*10^5$ mm$^{-2}$ at the detector position. With a detection efficiency of 1,5 % the number of detected neutrons is $1,2*10^4$ mm$^{-2}$. This might be considered a good starting point also for energy selective resonance imaging, provided the detection efficiency can be reproduced or even increased with the ULTRA8 and simultaneously upgrading of the accelerator to provide 20 uA beam in the pulsed mode is feasible.

### IV. SUMMARY

A description and first experimental results of a possible fast neutron radiography facility at PTB was presented. The neutron source has a good time structure for performing resonance radiography in a neutron beam of a broad spectral distribution. An increase by a factor of 10 in pulsed neutron intensity is desirable and, due to the construction of the cyclotron, this seems to be feasible either by upgrading the ion source or by implementing an external pulsed injection system. Three different types of detectors were tested, all showing specific advantages and drawbacks. The gas detector (FANGAS) offers high resolution (< 1mm, not shown here but measured with thermal neutrons [10]), fast timing, practically unlimited dynamic range and little gamma sensitivity. Drawbacks are the low detection efficiency. The optical detector with the counting image intensifier is not applicable in its present version. But its worthwhile to continue its investigation because of its potential as efficient neutron counting imaging device with superior timing and dynamic range capability as compared to its integrating counterpart..

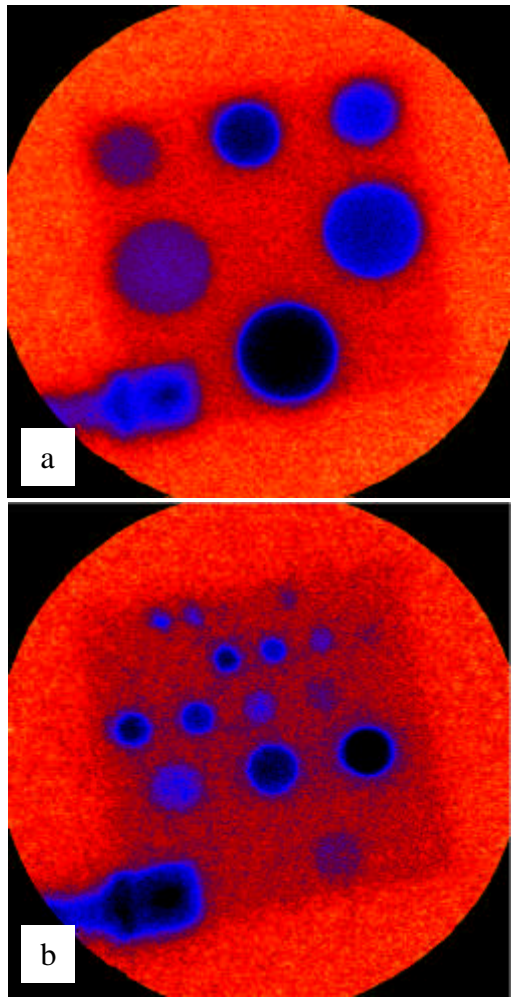

*Fig. 11: Radiography of (a) sample 1 and (b) sample 2 with full neutron spectrum obtained with OTIFANTI and CCD-camera readout. The images are summed up from 300 single exposures of 0,1 s each, i.e. the total exposure time is 30 s.*

The prospects of OTIFANTI with CCD readout depends on the future applicability of a fast framing camera. In our experiment the image intensified CCD system shows the largest neutron sensitivity, has an unlimited rate capability (which allows for much higher beam current as presently available) and will be therefore first choice for fast imaging applications where a small number of energy windows is sufficient.

ACKNOWLEDGMENT

We thank K. Tittelmeier, W. Heinemann and J. Hübner for their help during the preparation of these experiments. We also acknowledge the support of the accelerator staff at PTB in providing an excellent neutron beam for our experiments.